# Etched graphene single electron transistors on hexagonal boron nitride in high magnetic fields

A. Epping[1,2], S. Engels[1,2], C. Volk[1,2], K. Watanabe[3], T. Taniguchi[3], S. Trellenkamp[2], and C. Stampfer[1,2]

[1] *JARA-FIT and II. Institute of Physics, RWTH Aachen University, 52074 Aachen, Germany*
[2] *Peter Grünberg Institute (PGI-8/9), Forschungszentrum Jülich, 52425 Jülich, Germany*
[3] *National Institute for Materials Science, 1-1 Namiki, Tsukuba, 305-0044, Japan*

(Dated: November 19, 2013)

We report on the fabrication and electrical characterisation of etched graphene single electron transistors (SETs) of various sizes on hexagonal boron nitride (hBN) in high magnetic fields. The electronic transport measurements show a slight improvement compared to graphene SETs on $SiO_2$. In particular, SETs on hBN are more stable under the influence of perpendicular magnetic fields up to 9T in contrast to measurements reported on SETs on $SiO_2$. This result indicates a reduced surface disorder potential in SETs on hBN which might be an important step towards clean and more controllable graphene QDs.

**1 Introduction** Graphene is a very promising material for quantum dots (QDs) with potentially long spin relaxation times, due to its weak spin-orbit interaction [1, 2] and weak hyperfine coupling [3]. However, the lack of a band gap and the Klein tunneling phenomenon makes the confinement of electrons and the control of spin states challenging. Most experiments follow either the strategy of opening up a gap by etching graphene nanostructures and thus introducing a disorder-induced energy gap [4-10] or by gating bilayer graphene [11–16]. Though first results following the latter approach have been published, it is still at an early stage. Experiments on nanostructured graphene QDs have already allowed to demonstrate the electron-hole crossover [17], spin states [18] and charge relaxation times [19]. However, graphene nanostructures, e.g. graphene single electron transistors, on $SiO_2$ suffer from high disorder potential arising from the substrate and the edges making it hard to operate the SETs under the influence of high magnetic fields [20]. Due to its atomically smooth surface and graphene-like honeycomb lattice structure hexagonal boron nitride (hBN) is a promising candidate as a substrate for hosting graphene nanostructures [21]. It has been shown that the substrate induced disorder potential in graphene is indeed substantially lowered on hBN compared to $SiO_2$ [21, 22]. Graphene on $SiO_2$ suffers from charge puddles with diameters on the order of a few tens of nm [23], whereas the charge puddle size in graphene on hBN proved to be roughly one order of magnitude larger. So far the contribution of the edges to the overall disorder remains unknown.

In this article, we investigate nanostructured graphene single electron transistors on hBN with island sizes ranging from 100 to 300 nm under the influence of high magnetic fields. The sizes of the SETs are on the order of the expected charge puddle size of graphene on hBN. We performed low temperature electrical measurements at around 1.5 K in order to characterize our devices. The results are compared with measurements of similar devices on $SiO_2$. The analysis focuses on the magnetic field dependent transport through the SET. We show that the devices exhibit a stable and well defined single-dot behavior in magnetic fields up to 9 T. The results indicate that the disorder potential is lowered in graphene single electron transistors on hBN as compared to $SiO_2$, where a breaking of the SET into several SETs can be observed in high magnetic fields [20].

**2 Fabrication** As a first step hBN flakes are mechanically exfoliated onto highly doped silicon substrates with a 295 nm $SiO_2$ toplayer. Following the work of Dean et al. [21] individual mechanically exfoliated graphene flakes are transferred onto selected hBN flakes, which have a thickness of around 20-30 nm (see Fig. 1(a)). To structure the graphene flakes standard electron beam lithography (EBL) followed by reactive ion etching with an $Ar/O_2$ plasma is used. The etching is followed by an annealing step in ultra-high vacuum of $5\times10^{-8}$ mbar at 450 C° for 4 h. To contact the resulting graphene nanostructures a second EBL step followed by metal evaporation of Cr/Au is applied. To identify single-layer graphene, Raman spectroscopy measurements are performed. Fig. 1(e) shows a typical Raman spectrum of graphene on hBN. The signatures of the *LO* phonon of the hBN at 1365 cm$^{-1}$, the *G*-peak at 1584 cm$^{-1}$ and the *2D*-peak at 2680 cm$^{-1}$ with a full width half maximum (FWHM) of 25 cm$^{-1}$ shows the single-layer nature of the investigated graphene flake on the hBN substrate. A more detailed Raman spectroscopy study (see ref. [24]) of graphene on hBN shows the high quality of the fabrication process of our graphene on hBN. The Raman spectroscopy study indicates considerably lower doping fluctuations and a lower overall doping level compared to graphene on $SiO_2$. Figs. 1(b) shows a schematic illustration of an etched graphene SET on hBN. In Figs. 1(c)-(e) we show scanning force micrographs (SFM) of etched graphene SETs on hBN with different diameters. All devices have the same design, but differ in the size of the graphene island. The graphene islands are connected by narrow constrictions to source (S) and drain (D) leads. These graphene constrictions act as effective tunneling barriers. The chemical potential of the constrictions and the island can be individually tuned by the lateral graphene gates (see SGL, PG, SGR in Fig. 1(d)). The overall Fermi level can be adjusted by the underlying highly doped Si substrate acting as a back gate (BG).

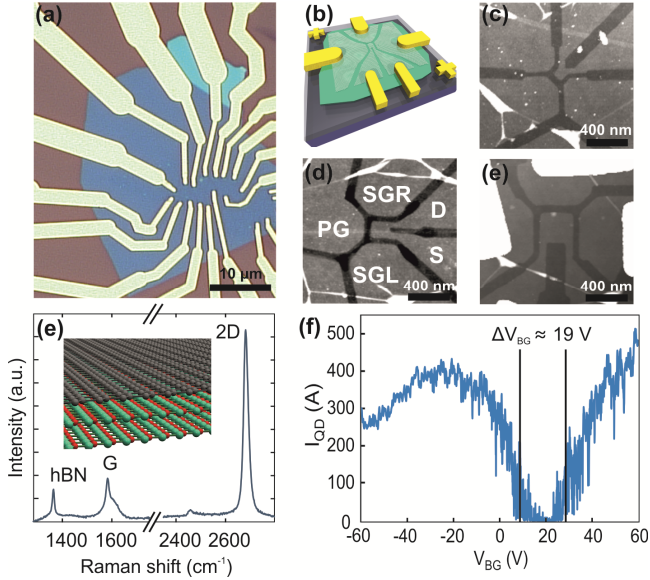

**Figure 1:** (a) Optical image of a graphene flake on hBN. (b) Schematic illustration of a graphene single electron transistor on hexagonal boron nitride (hBN). (c)-(e) atomic force micrographs of etched graphene SETs on hBN with different diameters $d$ ((c) $d = 110$ nm, (d) $d = 180$ nm and (e) $d = 300$ nm). (e) Raman spectrum of a representative graphene flake on hBN. (f) Back gate ($V_{BG}$) dependence of the current $I_{SD}$ through the graphene SET with $d = 180$ nm.

**3 Characterisation** In Fig. 1(f) and Fig. 2 we show low-temperature transport measurements ($T = 1.5$ K) performed on individual single electron transistors. Fig. 1(f) shows the source-drain current $I_{QD}$ as function of the back gate voltage $V_{BG}$ of nanostructured graphene on hBN with an island size of 180 nm (all side gate voltages are at 0 V) with a fixed bias of $V_{bias} = 300$ µV. Around $V_{BG} = 20$ V the current is mostly suppressed in a range of $\Delta V_{BG} = 19$ V (the so-called transport gap), which is in agreement with earlier studies on etched graphene SETs and QDs [7–9, 17, 18] and nanoribbons [4–6, 25–28] on SiO$_2$ and with etched graphene nanoribbons on hBN [29]. In order to investigate the influence of the hBN substrate on the properties of the graphene constrictions, large scale conductance measurements as function of the side gate voltages are performed. Fig. 2(a) shows the conductance of a 180 nm SET with 60 nm constrictions as function of the left and right side gate voltage. The back gate is set to a voltage inside the transport gap ($V_{BG} = 22$ V) and both side gates are swept from -15 V to 15 V while measuring the source-drain current. Transport through the individual graphene constrictions is mainly tuned by its closest side gate, visible by the relative lever arm of around $\alpha_{SGR/SGL} = 0.2$. The transport gap of both graphene constrictions divides the measurement in four areas of high conductance where both constrictions are transparent and a conductance on the order of 0.03 $e^2/h$ is observed. Fig. 2(b) shows a schematic illustration of the effective band structure of our devices, highlighting the two tunneling barriers leading to the four different transport configurations. By adjusting the potential in each constriction independently, it is possible to have either pure electron transport (NN), pure hole transport (PP) or a combination of both (NP or PN). In Figs. 2(c) and 2(d) finite bias measurements on a 110 nm SET with two 50 nm wide graphene constrictions are shown. The bias voltage is swept from –15 mV to 15 mV and the side gate voltage from –15 V to 15 V, while the other side gate is kept fixed. Both graphene constrictions exhibit a suppressed current over a large range of side gate voltage of $\Delta V_{gap}$ around 6 V, most likely arising from statistical Coulomb blockade [5, 30]. The gap in bias voltage is determined by the charging energy of the smallest charge puddle in the constriction resulting in an effective energy gap $E_g$. $E_g$ can be estimated from the finite bias measurements to be on the order of $E_{g,SGL} = 5$ meV and $E_{g,SGR} = 6$ meV. Compared to graphene nanodevices on SiO$_2$, one would expect a smaller effective energy gap on hBN due to larger charge puddles. However, the experimental results are in good agreement with the model derived for nanoribbons on SiO$_2$. According to Sols et al. [30] and Han et al. [4], the size of the effective energy gap is related to the width of the nanoribbon by $E_g = B^{-1} W^{-1} \exp(-CW)$ with $B = 0.001$ meV$^{-1}$nm$^{-1}$ and $C = 0.023$ nm$^{-1}$. Using these parameters for a 50 nm graphene nanoribbon translates in $E_g$ (50 nm) $\approx 6$ meV, which is in agreement with our results.

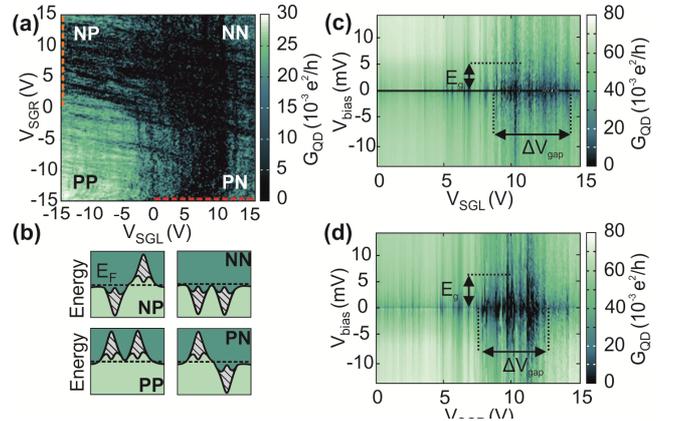

**Figure 2** (a) Conductance $G_{QD}$ as function of the side gate voltages $V_{SGR}$ and $V_{SGL}$ at $V_{BG} = 20$ V and $V_{bias} = 300$ µV. The constrictions can be separately tuned into the hole (P) and electron (N) transport regime. The transition regions are governed by the transport gap of each constriction (see schematic illustration (b)): Outside of the transport gaps, the constrictions are either both in an electron regime (NN), hole regime (PP) or a combination of both (NP/PN). (c)-(d) $G_{QD}$ as function of $V_{SGL}$ (c), $V_{SGR}$ (d) and $V_{bias}$ at $V_{SGR/SGL} = -15$ V. The transport gap and the effective energy gap of the constrictions are extracted to be $\Delta V_{SGR} \approx 5.5$ V / $\Delta V_{SGL} \approx 6$ V and $E_{g,SGR} = 6$ meV / $E_{g,SGL} = 5$ meV, respectively.

**3 Magnetotransport** After investigating the tunnel-tunneling barriers of the device, we now focus on quantum transport in high magnetic fields. Fig. 3(a) shows the con-conductance through a 110 nm single electron transistor while sweeping the back gate over the range of -55 V to 55 V and sweeping the magnetic field up to 8 T. Two distinct regions can be identified. The transport gap of the two graphene constrictions is visible around 27.5 V and extends over a range of around 11 V. The size of the gap is nearly independent of the magnetic field and can be converted to an energy of around 100 meV by using the energy dispersion and the lever arm of the back gate obtained in a Hall bar measurement. The position of the charge neutrality point is at $V_{QD} = 27.5$ V. A second regime is visible around zero back gate voltage, where resonances are observed, which move linearly with the magnetic field. The linear slope of these resonances indicates the formation of Landau levels. The magnetic field dependent resonances are not visible inside the transport gap, but only the shifting of the Coulomb resonances of the SET as function of magnetic field can be seen (Fig. 3(c)). The resonances indicating Landau levels are again visible outside the transport gap at high back gate voltages around 40 V. The resonances around zero back gate voltage originate most likely from the graphene leads and display a huge doping offset to the region of the graphene constrictions and the graphene island. Hence, it can be assumed that the etching process leads to an increase in doping in the region of the SET. The same behaviour has also been observed in our larger SETs. Due to the high number of localised states, which can be seen in Fig. 3(b), and multiple overlapping Landau level fans, which arise from differently doped regions of the leads, the zero Landau level is invisible. This makes it hard to really determine the positions of the Landau level fans and to distinguish between Landau levels and resonances originating from localised states. In addition, magnetic fields can strongly affect the stability of the formed SET. As shown by Güttinger et al. [20], SETs on SiO$_2$ tend to break apart when applying high magnetic fields. This can be related to the bulk disorder potential induced by the SiO$_2$ substrate. By applying a magnetic field, the electrons inside the graphene island start to be also magnetically confined. If the magnetic length $l_c = \sqrt{\hbar/eB}$ of the electrons on the graphene island is on the order of the disorder potential length scale, the electrons start to accumulate inside the charge puddles induced by the substrate. If the graphene island is bigger than the disorder potential length, the electrons can accumulate in different charge puddles leading to the formation of two or more individual single electron transistors. In graphene on SiO$_2$ the disorder potential length is estimated by Martin et al. [23] with surface potential measurements to around 30 nm. This length scale competes with the magnetic length of $l_c = $ 9 nm at 9 T, which can lead to the described breaking of the SET. As a result of the reduced bulk disorder on a hBN substrate, the breaking apart of SET is assumed not to occur. In Fig. 4 we compare Coulomb diamond measurements of the 300 nm SET at 0 T (Fig. 4(a)) and 9 T (Fig. 4(c)). Figs. 4(b) and 4(d) show corresponding charge stability diagrams. The charging energy of the SET at 0 T can be estimated to $E_c = 3$ meV leading to a dot diameter of roughly 700 nm. The relative lever arm is

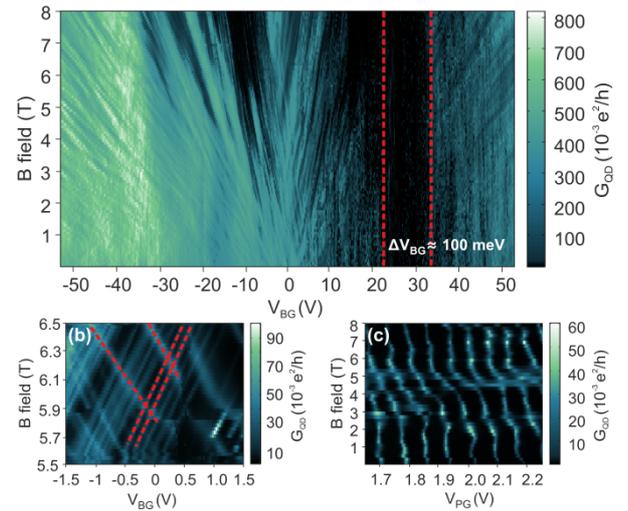

**Figure 3** (a) Conductivity $G_{QD}$ as function of the magnetic field and the back gate voltage at fixed $V_{bias} = 300$ μV (island size = 110 nm). Around zero voltage resonances occur, which scale linearly with the B field. The transport gap of the constrictions and the graphene island is offset by $V = 27.5$ V. The gap ranges from 22 V to 33 V. (b) High resolution measurement of (a) in the regime from 5.5 T to 6.5 T and from -1.5 V to 1.5 V in $V_{BG}$. Multiple resonances due to localised states are visible. (c) Conductance as function of $B$ field and $V_{PG}$ with back gate voltage fixed in the transport gap at $V_{BG} = 30$ V.

determined to be $\alpha_{PG/SGL} = 1.25$. Both, the charging energy and the relative lever arm allow the conclusion that the single electron transistor extends into the constrictions. However, it can be seen by the charge stability diagram (Fig. 4(d)), that the SET is not breaking apart at 9 T, although the conductance is more than one order of magnitude larger than at 0 T. This increase is caused by an increasing transparency of the tunnelling barriers. The same effect is also observed in the Coulomb diamond

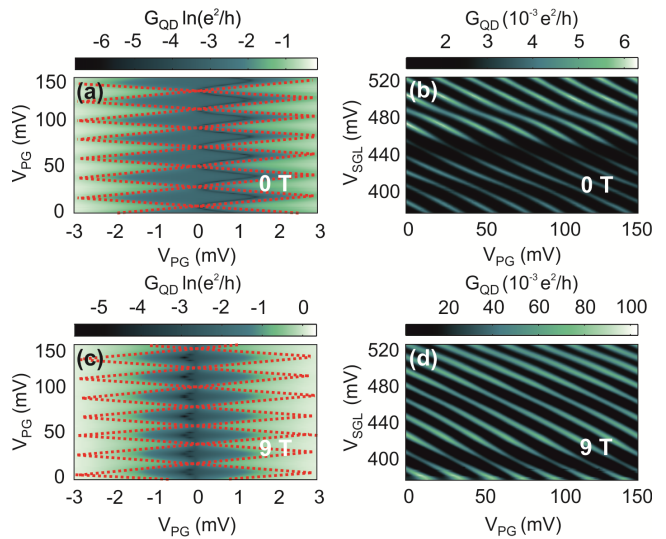

**Figure 4** (a) and (c) Conductance $G_{QD}$ of a single electron transistor on hBN with a diameter of $d = 300$ nm in dependence of $V_{bias}$ and $V_{PG}$ at a perpendicular magnetic field of (a) $B = 0$ T and (c) $B = 9$ T. Coulomb diamonds are visible with a B field independent addition energy of $E_{add} = 3$ meV and an increasing conductance at 9 T. (b) and (d) $G_{QD}$ in as function of $V_{PG}$ and $V_{SGL}$ for (b) $B = 0$ T and (d) $B = 9$ T. The charge stability diagram exhibits a stable single dot behavior independent of the magnetic field.

measurements at 9 T. Due to the conductance increase, the diamond shape is no longer visible above $V_{bias} \sim 2$ mV. Still, the lever arms and the spacing of the Coulomb resonances in *PG* voltage is very similar compared to 0 T ($\alpha_{PG/SGL,0T} = 1.25$ vs. $\alpha_{PG/SGL,9T} = 1.20$). This allows the conclusion that the SET is stable and well defined at 9 T over a large range of plunger gate voltage and occupies the same region on the graphene island compared to 0 T. The same behaviour has also been seen on a 180 nm SET. The observation of stable, large single electron transistors supports the assumption of a reduced bulk disorder potential due to the hBN.

**4 Summary** We present the characterization and analysis of electronic transport measurements at high magnetic fields performed on various etched single electron transistors (SETs) on hexagonal boron nitride (hBN). We show strong similarities to etched SETs and quantum dots fabricated on $SiO_2$. Magnetotransport measurements performed on samples of different island sizes show a doping mismatch between the lead regions and the region of the constriction/graphene island probably induced by the etching process. This effect can be identified by the magnetic field independent transport gap and B-field dependent resonances arising around zero back gate voltage. Each device shows this significant doping mismatch between the lead regions and the constriction / graphene island region on the order of $V = 27.5$ V. The resonances around $V_{BG} = 0$ can be attributed to the formation of Landau levels in the lead region, although the high number of localised states mask the zero Landau level. Furthermore we show that the 300 nm SET behaves like a stable single-dot, independent of the applied perpendicular magnetic field. We argue that these results can be explained by a decrease in the total disorder potential for larger single electron transistors on hBN where the substrate disorder might play a dominant role.

**Acknowledgements** We thank A. Steffen, H.-W. Wingens for help on sample fabrication and U. Wichmann for help and useful discussions on electronics. The support of the DFG (SPP-1459 and FOR-912) and the ERC are gratefully acknowledged.

**References**

[1] D. Huertas-Hernando, F. Guinea, A. Brataas, Phys. Rev. B **74**, 155426 (2006).
[2] H. Min, E. J. Hill, N. A. Sinitsyn, B. R. L. Sahu,Kleinman, A. H. MacDonald. Phys. Rev. B **74**, 165310 (2006).
[3] B. Trauzettel, D. V. Bulaev, D. Loss, G. Burkard. Nat. Phys. **3**, 192 (2007).
[4] M. Y. Han, B. Özyilmaz, Y. Zhang, and P. Kim, Phys. Rev. Lett. **98**, 206805 (2007).
[5] K. Todd, H.-T. Chou, S. Amasha, and D. Goldhaber-Gordon, Nano Lett. **9**, 416 (2009).
[6] C. Stampfer, J. Güttinger, S. Hellmüller, F. Molitor, K. Ensslin, and T. Ihn, Phys. Rev.Lett. **102**, 056403 (2009).
[7] S. Schnez, F. Molitor, C. Stampfer, J. Güttinger, I Shorubalko, T. Ihn, and K. Ensslin, Appl. Phys. Lett. **94**, 012107 (2009).
[8] X. L. Liu, D. Hug, L. Vandersypen, Nano Lett. **10**, 1623 (2010).
[9] S. Moriyama, D. Tsuya, E. Watanabe, S. Uji, M. Shimizu, T. Mori, T. Yamaguchi, and K. Ishibashi, Nano Lett. **9**, 2891 (2009).
[10] J. Cai, P. Ruffieux, R. Jaafar, Marco Bieri, T. Braun, S. Blankenburg, M. Muoth, A. P. Seitsonen, M. Saleh, X. Feng, K. Müllen, and R. Fasel, Nature **466**, 470 (2010).
[11] E. V. Castro, K. S. Novoselov, S. V. Morozov, N. M. R. Peres, J. M. B. L. dos Santos, J. Nilsson, , F. Guinea, A. K. Geim, A. H. C. Neto, Phys. Rev. Lett. **99**, 216802 (2007).
[12] J. B. Oostinga, H. B. Heersche, X. Liu, A. F. Morpurgo, and L. M. K. Vandersypen Nature Materials **7**, 151 (2007).
[13] T. Taychatanapat and P. Jarillo-Herrero Phys. Rev. Lett. **105**, 166601 (2010).
[14] J. Milton Pereira, P. Vasilopoulos, F. M. Peeters, Nano Lett. **7**, 946 (2007).
[15] M. T. Allen, J. Martin, and A. Yacoby, Nat. Com. **3**, 934 (2012).
[16] A. M. Goossens, S. C. M. Driessen, T. A. Baart, K. Watanabe, T. Taniguchi, and L. M. K. Vandersypen, Nano Lett. **12**, 4656-4660 (2012).
[17] J. Güttinger, C. Stampfer, F. Libisch, T. Frey, J. Burgdörfer, T. Ihn, and K. Ensslin, Phys. Rev. Lett. **103**, 046810 (2009).
[18] J. Güttinger, T. Frey, C. Stampfer, T. Ihn, and K. Ensslin, Phys. Rev. Lett. **105**, 116801 (2010).
[19] C. Volk, C. Neumann, S. Kazarski, S. Fringes, S. Engels, F. Haupt, A. Müller and C. Stampfer, Nat. Commun. **4**, 1753 (2013).
[20] J. Güttinger, C. Stampfer, T. Frey, T. Ihn and K. Ensslin, Nanoscale Res. Lett. **6**, 253 (2011).
[21] C. R. Dean, A. F. Young, I. Meric, C. Lee, L. Wang, S. Sorgenfrei, K. Watanabe, T. Taniguchi, P. Kim, K. L. Shepard and J. Hone, Nat. Nano. **5**, 722 (2010).
[22] J. Xue, J. Sanchez-Yamagishi, D. Bulmash, P. Jacquod, A. Deshpande, K. Watanabe, T. Taniguchi, P. Jarillo-Herrero and B. J. LeRoy, Nat. Mat. **10**, 282 (2011).
[23] J. Martin, N. Akerman, G. Ulbricht, T. Lohmann, J. H. Smet, K. von Klitzing, and A. Yacoby, Nat. Phys. **4**, 144 (2008).
[24] F. Forster, A. Molina-Sanchez, S. Engels, A. Epping, K. Watanabe, T. Taniguchi, L. Wirtz, and C. Stampfer, Phys. Rev. B **88** 085419 (2013).
[25] Z. Chen, Y.-M. Lin, M. Rooks and P. Avouris, Physica E **40**, 228, (2007).
[26] P. Gallagher, K. Todd and D. Goldhaber-Gordon, Phys. Rev. B **81**, 115409 (2010).
[27] M. Y. Han, J. C. Brant and P. Kim, Phys. Rev. Lett. **104**, 056801 (2010).
[28] B. Terrés, J. Dauber, C. Volk, S. Trellenkamp, U. Wichmann, C. Stampfer, Appl. Phys. Lett. **98**, 032109 (2011).
[29] D. Bischoff, T. Krähenmann, S. Dröscher, M. A. Gruner, C. Barraud, T. Ihn, and K. Ensslin, Appl. Phys. Lett. **101**, 203103 (2012).
[30] F. Sols, F. Guinea, and A. H. C. Neto, Phys. Rev. Lett., **99**, 166803 (2007).